\documentclass[aps,prd,showpacs,amssymb,superscriptaddress, notitlepage,floats,floatfix,nofootinbib,twocolumn]{revtex4-1}
\usepackage[utf8]{inputenc}
\usepackage{graphicx}
\usepackage{dcolumn}
\usepackage{bm}
\usepackage{float}
\usepackage{soul}
\usepackage{amsmath}
\usepackage{color}
\usepackage[dvipsnames]{xcolor}
\usepackage{hyperref}
\hypersetup{colorlinks=true, citecolor=MidnightBlue,linkcolor=CornflowerBlue, urlcolor=CornflowerBlue, linktocpage=true}
\usepackage{xfrac}
\usepackage{orcidlink}
\newcommand{\orcid}[1]{\href{https://orcid.org/#1}{\includegraphics[width=10pt]{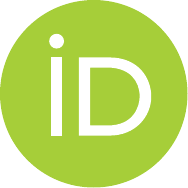}}}

\begin{document}

\title{Role of black hole quasinormal mode overtones for ringdown analysis}

\author{Peter James Nee\,\orcid{0000-0002-2362-5420}}
\email{peter.nee@aei.mpg.de}
\author{Sebastian H. V\"olkel\,\orcid{0000-0002-9432-7690}}
\author{Harald P. Pfeiffer\,\orcid{0000-0001-9288-519X}}
\affiliation{Max Planck Institute for Gravitational Physics (Albert Einstein Institute), D-14476 Potsdam, Germany}

\date{\today}

\begin{abstract}
Extracting quasinormal modes from compact binary mergers to perform black hole spectroscopy is one of the fundamental pillars in current and future strong-gravity tests. 
Among the most remarkable findings of recent works is that including a large number of overtones not only reduces the mismatch of the fitted ringdown but also allows one to extract black hole parameters from a ringdown analysis that goes well within the nonlinear merger part. 
At the same time, it is well understood that several details of the ringdown analysis have important consequences for the question of whether overtones are present or not, and subsequently, to what extent one can claim to perform black hole spectroscopy. 
To clarify and tackle some aspects of overtone fitting, we revisit the clearer problem of wave propagation in the scalar Regge-Wheeler and P\"oschl-Teller potentials. 
This setup, which is to some extent qualitatively very similar to the nonlinear merger-ringdown regime, indicates that using even an approximate model for the overtones yields an improved extraction of the black hole mass at early ringdown times. 
We find that the relevant parameter is the number of included modes rather than using the correct model for the overtones themselves. 
These results show that some standard tests for verifying the physical contribution of an overtone to a waveform can be misleading, and that even in the linear case it can be difficult to distinguish the presence of an excited mode from the fitting of non-QNM effects.
\end{abstract}

\maketitle

\section{Introduction}\label{int}

The ongoing success of the LIGO-Virgo-KAGRA Collaboration in measuring gravitational waves from binary black hole mergers finally allows us to probe the strong and dynamical field of general relativity (GR) \cite{LIGOScientific:2016aoc,LIGOScientific:2018mvr,LIGOScientific:2020ibl,LIGOScientific:2021djp}. 
Among the most promising and exciting challenges ahead lies black hole spectroscopy \cite{Detweiler:1980gk,Dreyer:2003bv,Berti:2005ys}. 
It allows for a ``fingerprint analysis'' of the final object by measuring its relaxation in terms of the emission of characteristic quasinormal modes (QNMs) within linear perturbation theory \cite{Regge:1957td,Zerilli:1970se,Teukolsky:1973ha}, after the two initial objects merged, ending a very long inspiral phase. 
The spectrum of QNMs in GR is fully characterized by the mass and spin of the final Kerr black hole \cite{Kerr:1963ud}, if the assumptions of the no-hair theorem are fulfilled \cite{Carter:1971zc,Robinson:1975bv}. 
Therefore, any measurement of frequencies and damping times that are not in agreement with this prediction points toward fundamental violations in our current understanding of black holes and compact objects. 

Several recent discussions in the literature motivate us to carefully revisit some basic paradigms and concepts about extracting QNMs from simulations and observations, as they are the pillar of performing black hole spectroscopy successfully. 
The two most relevant aspects in the context of this work are concerning the extraction of QNMs from simulations (which can, in principle, be as accurate as one desires) and from observations (which strongly depend on detector noise). 
On the observational side, there have been several works that report the robust extraction of fundamental modes \cite{LIGOScientific:2016lio,Carullo:2019flw,LIGOScientific:2020tif,LIGOScientific:2021sio}. 
Moreover, there is also the claim that the first overtone has been measured \cite{Isi:2019aib}, but the debate regarding the robustness of this claim is ongoing \cite{Cotesta:2022pci,Finch:2022ynt,Isi:2022mhy,Capano:2022zqm}. 
Because the disagreement involves, to a large extent, the correct modeling of the noise and the detector, future observations should resolve such problems \cite{Berti:2007zu,Baibhav:2017jhs,Giesler:2019uxc,Ota:2019bzl,Bhagwat:2019dtm,Bhagwat:2019bwv,Cook:2020otn,CalderonBustillo:2020rmh,JimenezForteza:2020cve,MaganaZertuche:2021syq}. 

On the theoretical/numerical side, a key challenge is quantifying the time after which the nonlinear merger can be described accurately by linear perturbation theory in terms of a superposition of QNMs. 
It is clear that this also plays a central role when analyzing data from the ringdown alone. 
Early studies about the significance of QNMs in the time domain in linear theory can be found in Refs.~\cite{Nollert:1996rf,Nollert:1998ys} and results based on the analysis of numerical relativity simulations can be found in Refs.~\cite{Buonanno:2006ui,London:2014cma}. 
The importance of nonlinearity has been further challenged by recent works demonstrating that using a large number of QNM overtones is enough to describe binary black hole merger events even around the peak of the strain, see Ref.~\cite{Giesler:2019uxc} and Refs.~\cite{Mourier:2020mwa,Forteza:2021wfq}. 
These findings seem to allow QNMs to capture earlier times of the merger-ringdown regime than is commonly expected. 
While the prompt reply to such claims might be possible overfitting, the authors have carried out several tests, e.g., extracting the injected black hole's mass or modifying the QNM spectrum in the analysis, which make their conclusions very strong. 
Somehow challenging this claim are recent works that address the question of whether nonlinear features in numerical relativity simulations can be robustly constrained when analyzing the early ringdown, see Refs.~\cite{Sberna:2021eui,Cheung:2022rbm,Mitman:2022qdl}. 
In fact, in these works it is argued that such nonlinear effects could even be more relevant than overtones, which would be a dramatic change in the standard perception and application of black hole perturbation theory. 

While nonlinear effects or the starting time of the linear regime are within the realm of numerical relativity, similar problems for QNM extraction are well-known from linear perturbation theory and deserve some review as they may be overlooked. 
The first aspect is to spell out that even the linear regime of a ringdown is in general not completely described by a superposition of QNMs, e.g., in contrast to linear oscillations of a string. 
This is due to QNMs not forming a complete basis, such that the initial data evolved is not completely described by them.  A second aspect is the presence of a late-time tail (Price tails \cite{Price:1971fb,Price:1972pw}), which eventually becomes the dominant feature. 
Thus, even if one would know when the full NR waveform is linear, one could not be sure that it is indeed well described by a superposition of QNMs alone. 
Fitting those modes in an agnostic way may thus lead to biased results. 

One popular toy model in the context of black hole QNMs is to consider the P\"oschl-Teller potential~\cite{Mashhoon:1982im,Ferrari:1984ozr,Ferrari:1984zz} as an approximation for the potential that appears in the perturbation equations. 
It has the well-known advantage that the spectrum of QNMs is known analytically and can be used as an approximation of those for the Regge-Wheeler, Zerilli, or other similar potentials. 
In the context of this work, it also has the interesting feature that there are no late-time tails when studying the time domain problem \cite{Beyer:1998nu}, which limits extraction of the fundamental mode at late times to just numerical errors. 
In the same work, it was also shown that this allows one to represent certain types of $\mathcal{C}^\infty$ initial data as a series of QNMs at late enough times, which is in contrast to the Regge-Wheeler potential.

The main ingredients of our work are as follows. 
First, we produce time-domain waveforms obtained by scattering Gaussian wave packets with the P\"oschl-Teller potential or the scalar Regge-Wheeler potential (hereby referred to as the GR potential). 
Second, we deploy a fitting scheme to extract QNMs using a certain model, starting time, and length of the extracted waveform. 
As for models, we use two different approaches, an agnostic one, and a theory-specific one. 
In the agnostic model each overtone is fitted with independent amplitude, phase, frequency, and damping time. 
In the theory-specific model, we consider either the P\"oschl-Teller QNM spectrum or the scalar Regge-Wheeler QNM spectrum (similarly, referred to as the GR QNM spectrum). 
As such, all frequencies and damping times are controlled by only one parameter, the mass, while the set of amplitudes and phases is fitted independently. 

We are able to explicitly demonstrate that including overtones in the theory-specific models allows one to better estimate the black hole mass at earlier starting times, even if the wrong model is being used for the analysis. 
More specifically, the relative error of the extracted mass as a function of the starting time at early times is remarkably similar, depending mainly on the number of overtones included and only mildly on the model itself. 
This raises several questions to be explored in future work; when using more overtones for the fitting, is it possible that one rather improves the correct fundamental mode fit by ``fitting away'' the initial data traces, thus obtaining the mass with more accuracy at earlier times? 
Why does it seem that including overtones, which are more sensitive to changes in the model (particularly their spectral stability), can improve the extraction of physical parameters (in our case, the mass), even when incorrect overtones are used?

This article is structured as follows. 
In Sec.~\ref{met} we outline the methods being used to generate our waveforms and statistical methods to analyze them. 
The application and results are then provided in Sec.~\ref{app}. 
Finally, we summarize in Sec.~\ref{con}. 
Throughout this work we use units in which $G=c=1$. 

\section{Methods}\label{met}

For an introduction to QNMs we refer the interested reader to Refs.~\cite{Kokkotas:1999bd,Berti:2009kk} for classic reviews and Refs.~\cite{Maggiore:2007ulw,Maggiore:2018sht} for comprehensive text books. 
Different types of field perturbations around the Schwarzschild black hole can be written in the form of a master equation,
\begin{align} \label{Mod_Wave}
\frac{\text{d}^2}{\text{d}t^2} \psi(t,x) - \frac{\text{d}^2}{\text{d}x^2} \psi(t,x) + V(x) \psi(t,x) = 0,
\end{align}
where $x$ is the tortoise coordinate;
\begin{align}
x = r + 2M \ln\left(\frac{r}{2M} - 1 \right).
\end{align} 
Here $V(x)$ is an effective potential that describes a barrier with a maximum located approximately around the light ring $3M$.  
In the test scalar field case ($\square \phi = 0$) the potential is given by
\begin{align}\label{pot_GR}
V_\text{GR}(r) = \left(1-\frac{2M}{r}\right) \left(\frac{l(l+1)}{r^2} + \frac{2M}{r^3} \right).
\end{align}
Note that the gravitational case, which can be split into axial and polar perturbations of the metric, yields qualitatively similar potentials known as the Regge-Wheeler and Zerilli potentials. 

The aforementioned P\"oschl-Teller potential is given by
\begin{align}\label{pot_PT}
V_\text{PT}(x) = \frac{V_0}{\cosh^2(\alpha (x - x_{0}))},
\end{align}
where $V_0$, $\alpha$, and $x_{0}$ are chosen such that
the maximum of the P\"oschl-Teller potential coincides with the one of the studied potential (here the GR potential), as well as their second derivatives in the tortoise coordinate (for explicit calculations, see Refs.~\cite{Ferrari:1984ozr,Ferrari:1984zz}). 
A Fourier decomposition of Eq.~\eqref{Mod_Wave} and suitable boundary conditions then leads to the QNMs as a set of complex frequencies; but in the following we are interested in the evolution of initial data in the time domain in the spirit of Vishveshwara's pioneering analysis \cite{Vishveshwara:1970zz}. 

In this work we solve Eq.~\eqref{Mod_Wave} numerically via a finite difference scheme (in particular, a central in time and central in space scheme):
\begin{multline} \label{CTCS}
\psi_{j}^{i} = 2\psi_{j}^{i-1}-\psi_{j}^{i-2}+\frac{\Delta t^2}{\Delta x^2}\left(\psi^{i-1}_{j+1}-2\psi^{i-1}_{j}+\psi^{i-1}_{j-1}\right)\\-\Delta t^{2}\,\psi_{j}^{i-1}V_{j}\text{.}
\end{multline}
Here $\psi^{i}_{j} = \psi(t_{i}, x_{j})$, $V_{j} = V(x_{j})$ and $\Delta t$ and $\Delta x$ are the temporal and spatial resolutions respectively. 
Our initial data consists of incoming Gaussian wave packets,
\begin{equation}\label{initialdata}
\psi(0,x) = Ae^{-\frac{(x - 30M)^{2}}{2d^{2}}},\quad\psi_{t}(0,x) = \psi_{x}(0,x),
\end{equation}
with an amplitude $A$ comparable to the maximum of the potential and widths $d$ comparable to the width of the potential.
Outgoing boundary conditions are imposed (although boundaries are chosen at a distance such that possible reflections will not contaminate the observed waveforms). 
We record the waveform $\psi(t)=\psi(t,R)$ at extraction radius $R$. 
We perform convergence tests for the code, vary the extraction radius $R$, consider waveforms generated from different initial data with different widths $d$, and also the impact of temporal-spatial resolution and length of different waveforms for our analysis. We found that these parameters did not qualitatively alter the results presented below.
Figure~\ref{waveforms} shows the waveforms generated in this way, which we use in the following analysis. 
\begin{figure}
\centering
\includegraphics[width=1.0\linewidth]{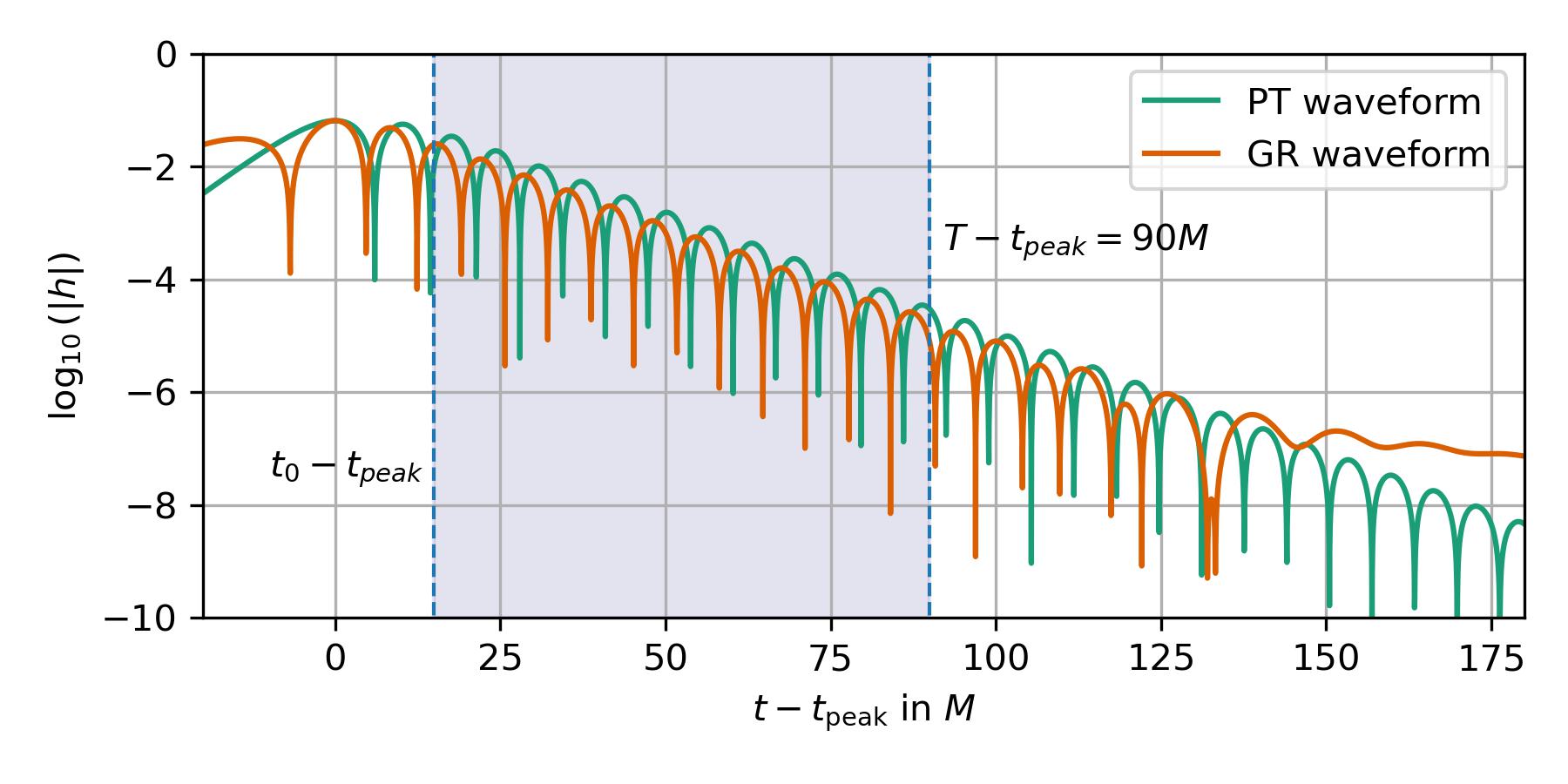}
\caption{Waveforms generated from the same Gaussian initial data evolved in the P\"oschl-Teller potential (green) and GR potential (orange). The shaded area indicates a typical fit interval, starting at $t_{0}-t_\mathrm{peak}$, and ending at $T-t_\mathrm{peak}$. Note that the time has been shifted to align both waveforms at their respective peak time $t_\mathrm{peak}$, which introduced a relative shift with respect to the simulation time (not shown).}
\label{waveforms}
\end{figure}

The extraction of QNMs is handled by fitting the numerical $\psi(t)$ in different ways as follows. 
In all models we first choose a starting time for the fit $t_{0}$, and a final time $T$.
In the agnostic model (AG) we then assume that the entire signal after the starting time can be written as 
\begin{align}
\psi_\text{AG}(t-t_\mathrm{peak}) = &\sum_{n=0}^{N-1} A^\text{AG}_n \exp\Big( - \omega^\text{AG}_{i,n} (t-t_\mathrm{peak}) \Big) \nonumber
\\
&\quad\times  \sin\Big(\omega^\text{AG}_{r,n} (t-t_\mathrm{peak}) + \phi^\text{AG}_n \Big),
\end{align}
where $t_\mathrm{peak}$ is defined as the maximum value of $|\psi(t)|$. Here each mode $n$ is characterized by an independent amplitude $A^\text{AG}_n$, phase $\phi^\text{AG}_n$, and complex mode frequency $\omega^\text{AG}_n = \omega^\text{AG}_{r,n} + i \omega^\text{AG}_{i,n}$, for a total of $4N$ fitted parameters.

For the theory-specific model the waveform is given by
\begin{align}
\psi_\text{TS}(t-t_\mathrm{peak}) = &\sum_{n=0}^{N-1} A^\text{TS}_n \exp\Big( - \omega^\text{TS}_{i,n}(M) (t-t_\mathrm{peak}) \Big)  \nonumber
\\ 
&\quad\times\sin\Big(\omega^\text{TS}_{r,n}(M) (t -t_\mathrm{peak}) +  \phi^\text{TS}_n \Big),
\end{align}
where the theory-specific model (TS) either stands for GR or P\"oschl-Teller (PT). 
Here the free parameters for a given mode $n$ are the amplitude $A^\text{TS}_n$ and phase $\phi^\text{TS}_n$. Furthermore, the mass $M$ is also a fitted parameter, which uniquely determines the complex mode frequencies for all $n$; hence, the GR/PT models have $2N+1$ fitted parameters.  For the GR potential we use the publicly available data for the QNM spectrum provided in Refs.~\cite{Berti:2005ys,Berti:2009kk}, and use the fact that $M \omega$ is constant. 
For the theory-specific model of the P\"oschl-Teller potential the QNM spectrum $\omega^\text{PT}$ can be obtained analytically as \cite{Mashhoon:1982im,Ferrari:1984ozr,Ferrari:1984zz}
\begin{align}
\omega^\text{PT} = \left(V_0 - \frac{\alpha^2}{4} \right)^{1/2} + i \alpha \left( n + \frac{1}{2} \right).
\end{align}

The best-fit values of a given model are obtained by using the \textsc{Python} optimize curve fit library. 
To perform these fits one has to specify the prior ranges for the fitted parameters. 
We found that for sensible choices such that the fit does not survey too large a range, our results were agnostic to this choice.  Furthermore, our recovered best-fit parameters never rail against the boundaries of the priors.
To compare our results with those of Ref.~\cite{Giesler:2019uxc} we define a similar mismatch function
\begin{align}
\mathcal{M}(h_1, h_2) = 1 - \frac{\left<h_1, h_2 \right>}{\sqrt{\left<h_1, h_1 \right> \left<h_2, h_2 \right>}},
\end{align}
where
\begin{align}
\left<h_1, h_2 \right> = \int_{t_0}^{T} h_1(t) h_2(t)\text{d}t.
\end{align}

Here $t_0$ and $T$ have the same meaning as introduced earlier, and we have omitted a customary complex conjugation on $h_2$, as all of our waveforms are real. 
To be confident that curve fitting does not get stuck in
a local minimum, we repeated the optimization 50 times with random initial parameters within reasonable prior ranges and selected the parameters yielding the smallest mismatch. This was also repeated for different numbers of repetitions, yielding consistent results. To be close to the setup of Ref.~\cite{Giesler:2019uxc} we also set $T=90 M + t_\mathrm{peak}$, although in their work it refers to the total mass of the system and in our case only to the remnant mass.
The dependence of the overall results on the choice of $T$ had been explored by treating it as a free parameter. 
Besides at very late times, when the tail dominates over the QNM contributions, the precise choice of $T$ does not play a significant role, as long as at least a few QNM oscillation times are captured.
Finally, the relative error of the black hole mass is defined as
\begin{align}
\delta M = \frac{|M_\text{rec}-M_\text{inj}|}{M_\text{inj}},
\end{align}
where $M_\text{rec}$ is the reconstructed mass and $M_\text{inj}$ the injected one, the mass-parameter entering the potential $V$ in Eqs.~(\ref{pot_GR}) and~(\ref{pot_PT}) used in our numerical solutions.

\section{Application and Results}\label{app}

As a first application we study the QNM fitting of Gaussian wave packets scattered at the P\"oschl-Teller potential.  We fit the numerical waveform with the AG, PT and GR models at different order $N$, and compute the mismatches between the original data and each fit. This procedure is repeated for many different starting times $t_0$ of the fit interval.

\begin{figure}
\includegraphics[width=1.0\linewidth]{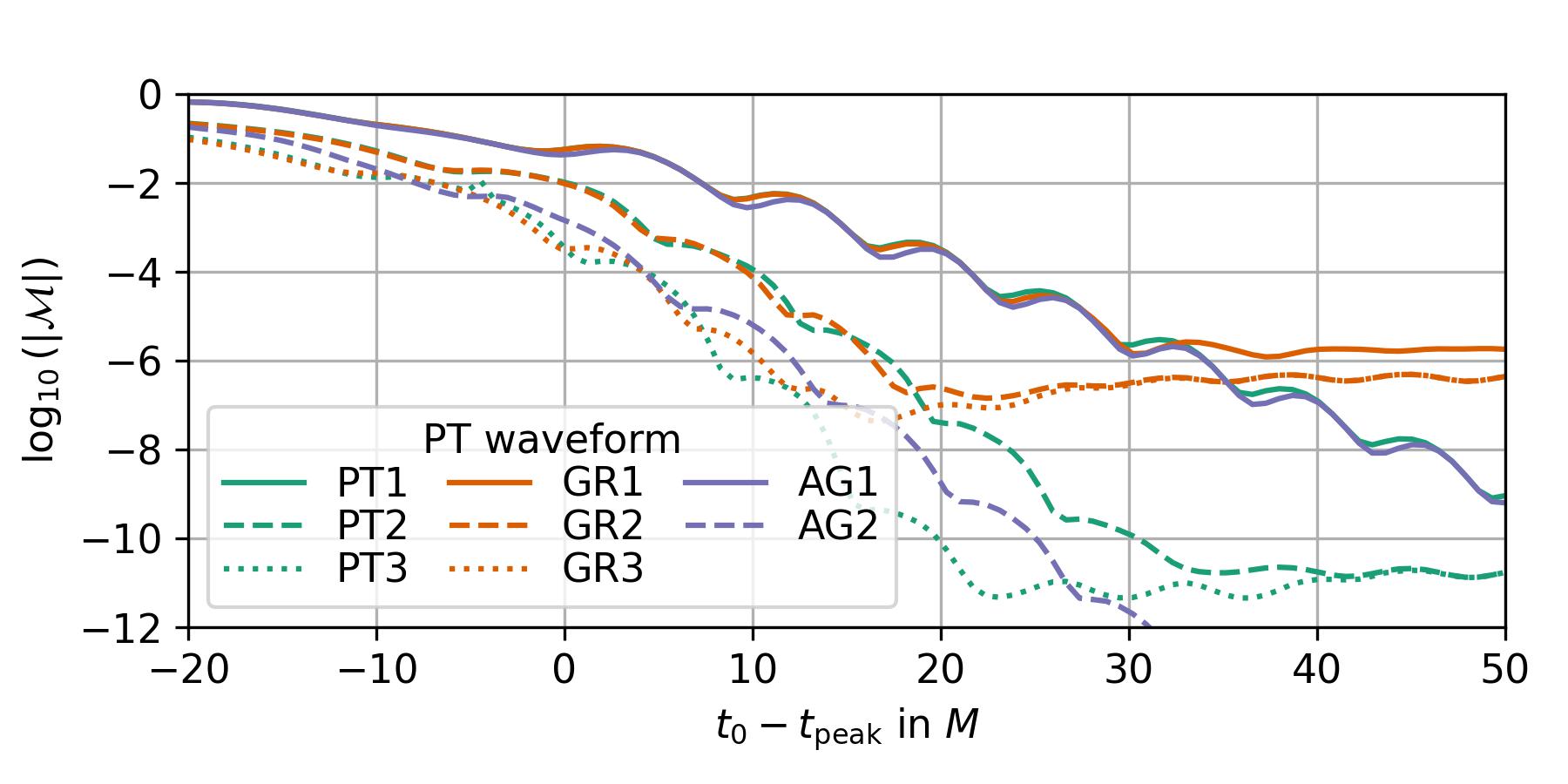}
\includegraphics[width=1.0\linewidth]{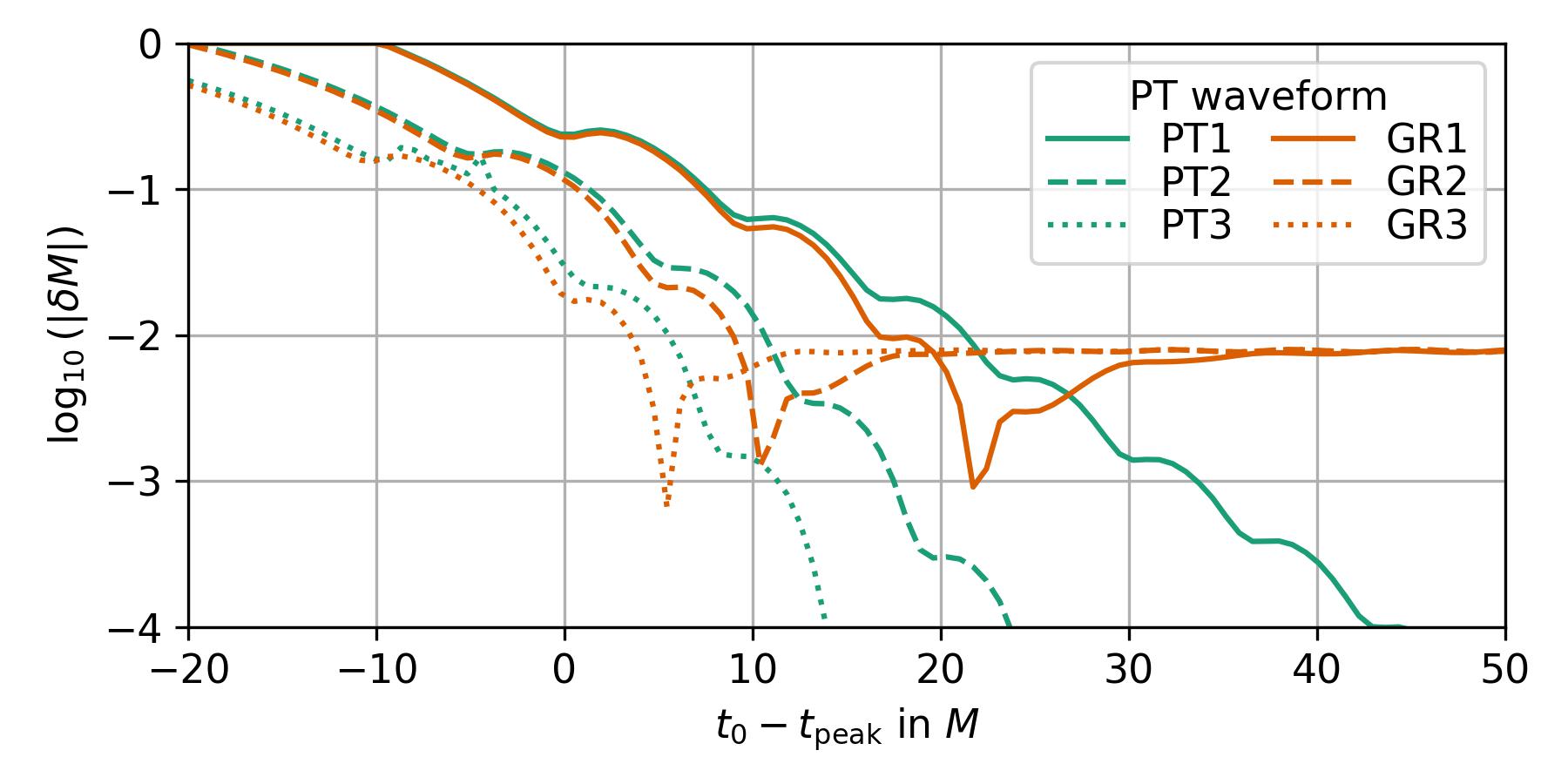}
\caption{
Results for a waveform generated using the P\"oschl-Teller potential. 
We apply the PT and GR model with $N=1\ldots3$ modes, as well as the agnostic model with $N=1$ or $N=2$ modes. 
\textbf{Top panel}: mismatch $\mathcal{M}$ as a function of starting time of the fit. 
\textbf{Bottom panel:} relative error $\delta M$ of the recovered black hole mass as a function of starting time of the fit. }
\label{PT_fig}
\end{figure}

Our results are summarized in Fig.~\ref{PT_fig}. The top panel shows the mismatch as a function of the beginning of the fit interval.
All three $N=1$ models 
yield similar results for the mismatch after the peak of the signal throughout most of the ringdown.  
Furthermore, it is remarkable that for small $t_0-t_{\rm peak}$ the two theoretical models at the same $N$ perform similarly.
Only at later times the (correct) PT model outperforms the GR model. 
Overall the GR models reach mismatches $\sim10^{-6}$, while the PT models reach $\sim10^{-8}$. 
Note that at very late times the agnostic model AG2 yields the smallest mismatch and even outperforms the theoretical models with multiple modes. 
We expect that numerical inaccuracies on such a small level are better described by the AG2 model due to its larger number of free parameters and independent QNMs. 

The bottom panel of Fig.~\ref{PT_fig} shows the relative error of the mass $\delta M$, as a function of the corresponding starting times. 
Since one cannot extract a mass from the agnostic models, we only compare the theory-specific models. 
As was the case for the mismatch, it is evident that the dependence at earlier times lies mainly in the number of modes used, and less so in the specific model being used. 
Somewhat surprisingly, for the same number of modes and early beginning of the fitting interval ($0\lesssim t_0-t_{\rm peak}\lesssim 10\ldots 20\,M$, depending on $N$), the GR model recovers the mass slightly better. 
Then, the number of used modes determines at what time the GR model plateaus toward a value of around $10^{-2}$. 
This is expected, because the fundamental modes of the two models agree with each other only to percent level. 

\begin{figure}
\includegraphics[width=1.0\linewidth]{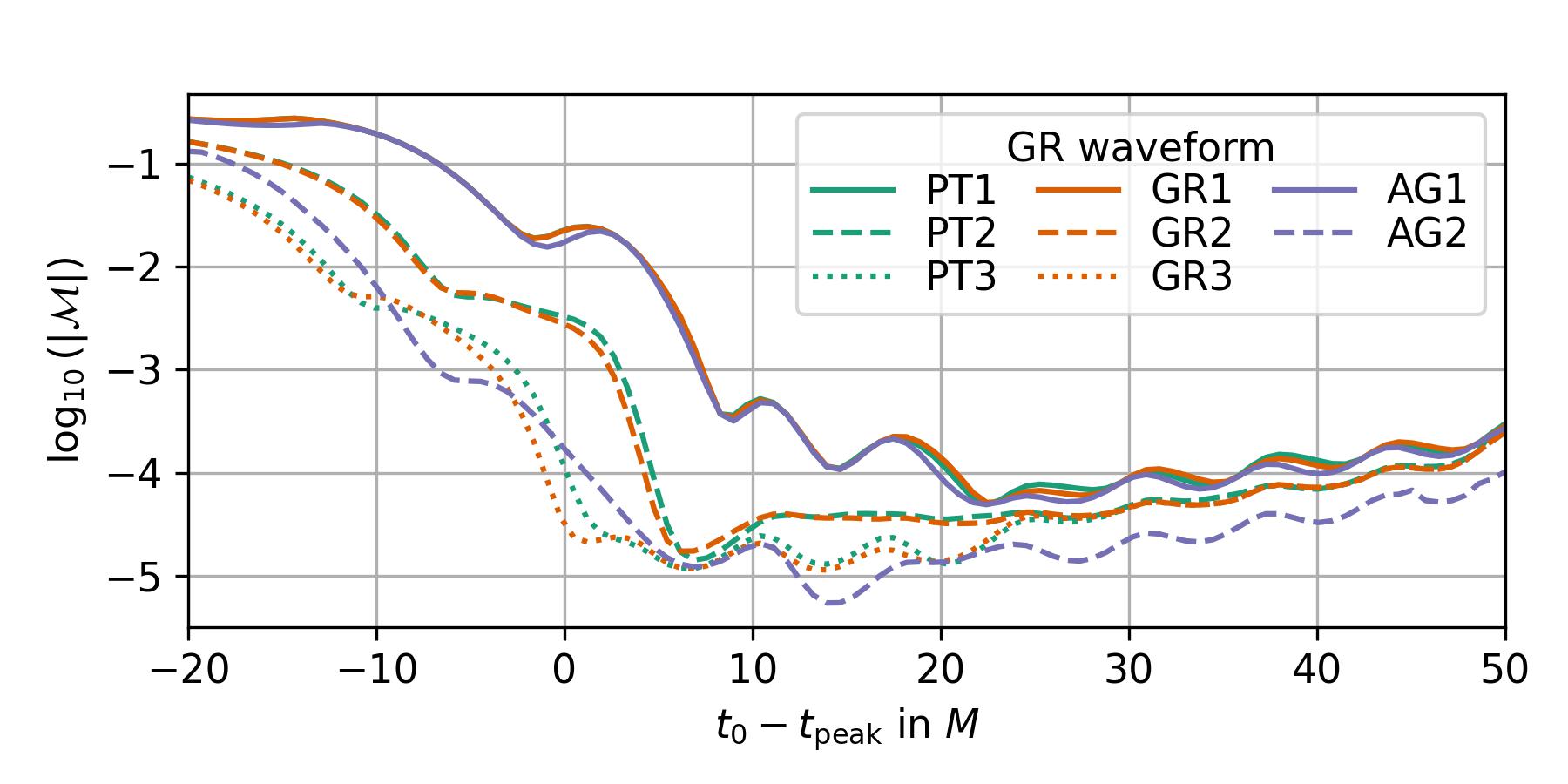}
\includegraphics[width=1.0\linewidth]{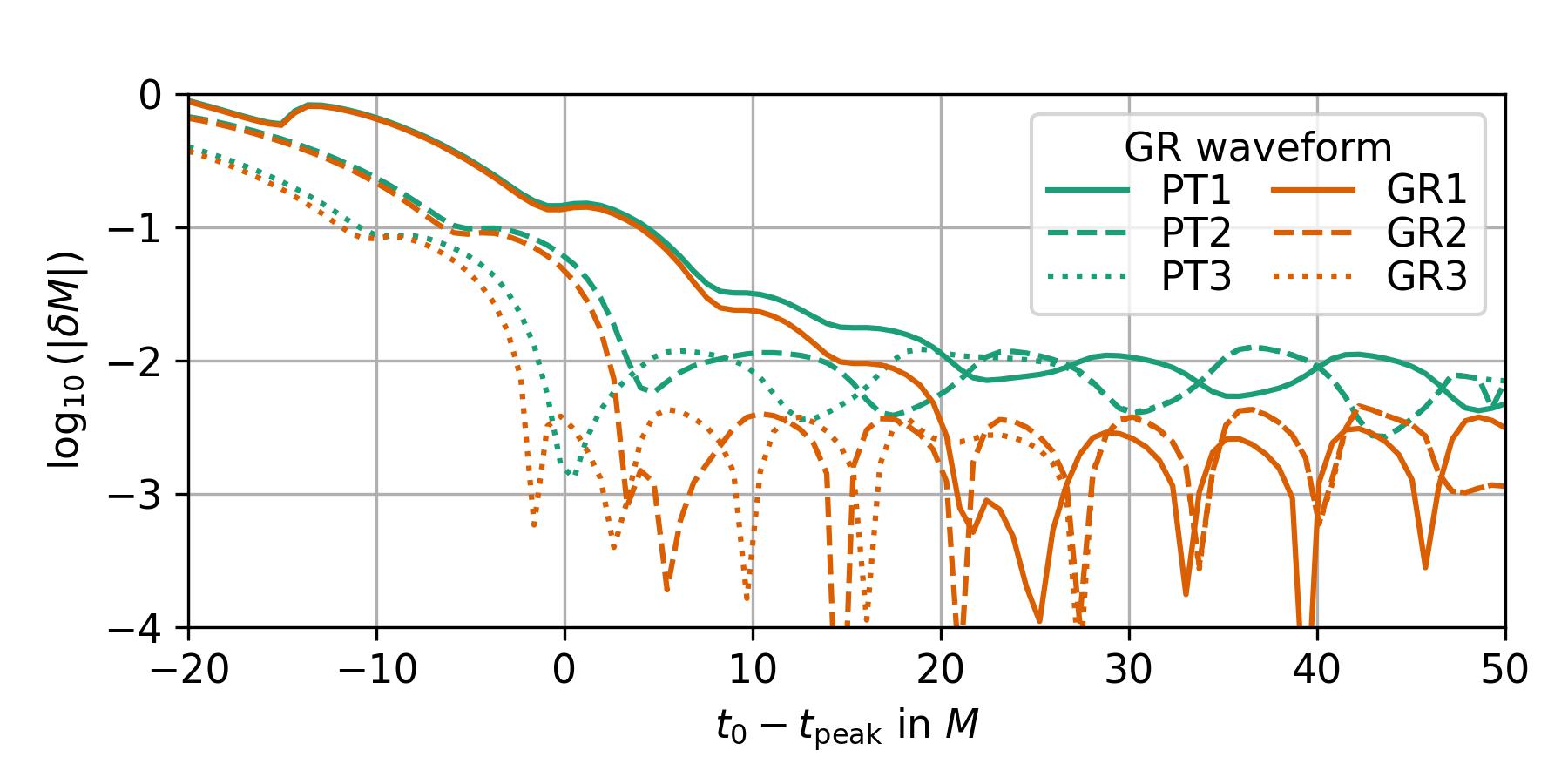}
\caption{Results for a waveform generated using the GR potential. 
We apply the PT and GR model with $N=1,2,3$ modes, as well as the agnostic model with $N=1$ or $N=2$ modes. 
\textbf{Top panel}: mismatch $\mathcal{M}$ as a function of starting time minus peak time of the waveform. 
\textbf{Bottom panel:} relative error $\delta M$ of the black hole mass as a function of starting time minus peak time of the waveform. 
At $t_0-t_{\rm peak}\gtrsim 10\ldots 20\,M$, the fits are limited by waveform tails.}
\label{GR_fig}
\end{figure}

As a second application, we repeat the previous analysis using a waveform produced with the GR potential.
Our results are summarized in Fig.~\ref{GR_fig} revealing similarities
as well as differences to Fig.~\ref{PT_fig}, as we will now discuss:
Regarding the mismatch we find a very similar behavior until around $t_0 - t_\mathrm{peak} \sim 10\ldots 20 M$. 
For later starting times of the fit, the mismatches of all models are comparable ($\mathcal{M}\sim 10^{-5}$) and even increase slowly. 
This is a clear indication of tail contributions, which limits the validity of the QNM expansion of the waveform. 

We now turn to the recovery of the mass in the PT and GR fits, shown in the lower panel of Fig.~\ref{GR_fig}.  For early start times, PT and GR models at the same order $N$ recover the mass comparably well.  The PT models then level off at a relative error $\delta M\sim 10^{-2}$ owing to the different fundamental mode frequency of the PT model, compared to the analyzed GR waveform.  The GR models recover the mass about 1 order of magnitude better.  However, despite the GR model employing the correct frequencies of the QNM modes for the analyzed GR waveform, mass recovery levels off at a few $\times 10^{-3}$, presumably due to the presence of waveform tails. 

\begin{figure*}
\includegraphics[width=1.0\linewidth]{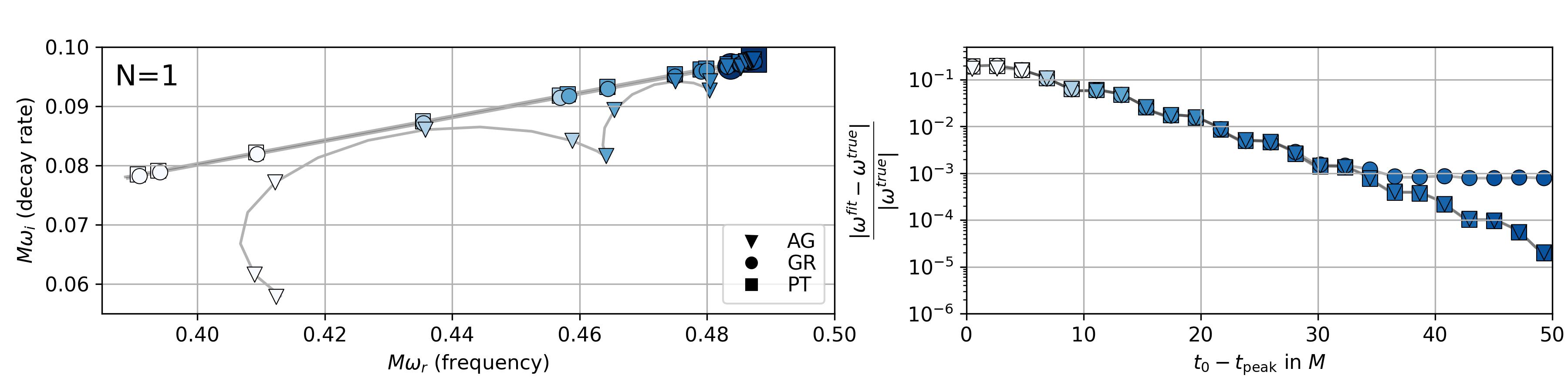}
\includegraphics[width=1.0\linewidth]{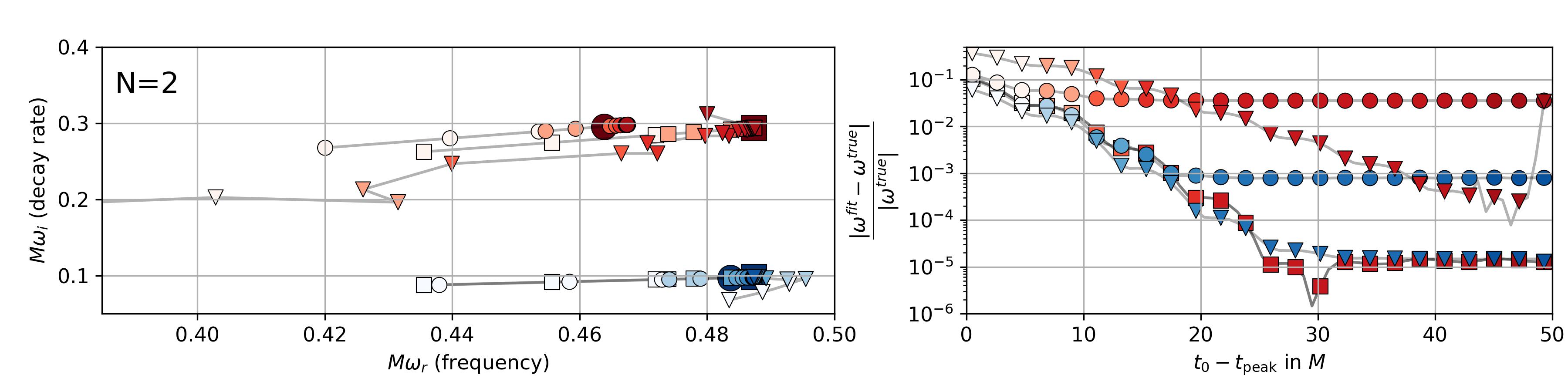}
\includegraphics[width=1.0\linewidth]{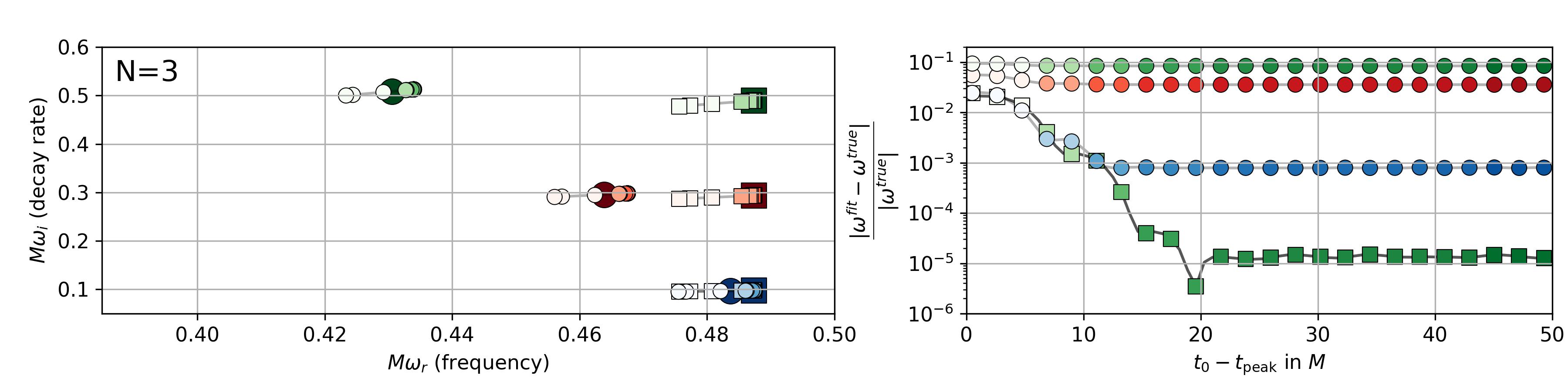}
\caption{
Analysis of a PT waveform using different fitting models. The \textbf{top left panel} shows the QNM frequencies recovered when fitting only one mode ($N=1$). The small markers indicate the recovered QNM frequencies for fits that start at different $t-t_\text{peak}$, as indicated by the shading along with the horizontal axis of the \textbf{top right panel}. The vertical axis of the top right panel shows the relative difference of the recovered QNM frequency to the fundamental mode of the PT waveform. This fundamental mode is also plotted with the large square in the top left panel. The \textbf{middle} panels and the \textbf{bottom panels} show the analogous results when fitting $N=2$ and $N=3$ modes, respectively, with the first overtone colored red and the second overtone colored green. For reference, the left panels also indicate the QNM frequencies of a GR waveform as large circles. Note that because the PT QNM frequencies are all proportional to the mass (which is fitted), all the PT QNM frequencies have the same relative errors and lie on top of each other in the right panels.}
\label{PT_qnm}
\end{figure*}

\begin{figure*}
\includegraphics[width=1.0\linewidth]{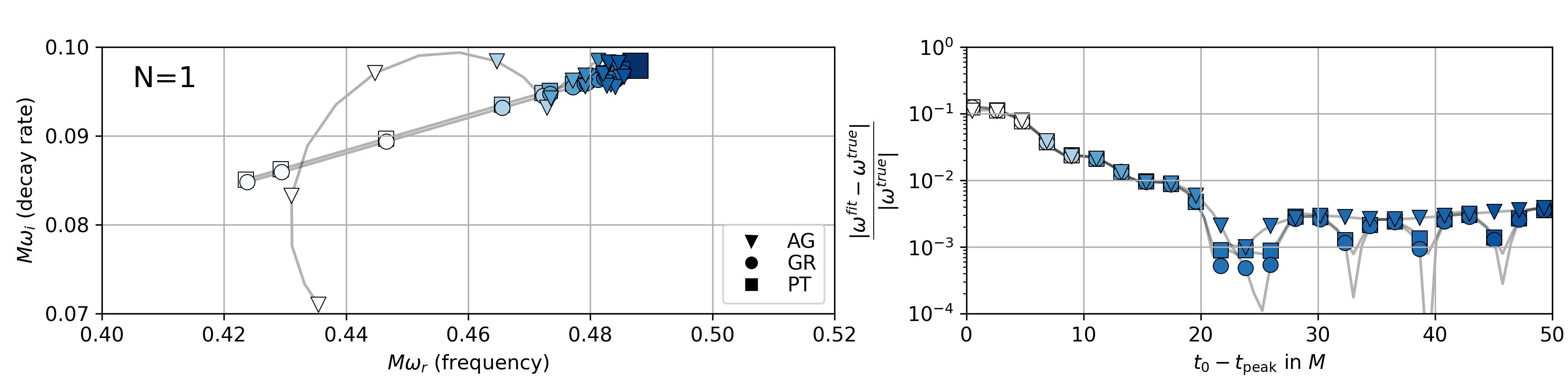}
\includegraphics[width=1.0\linewidth]{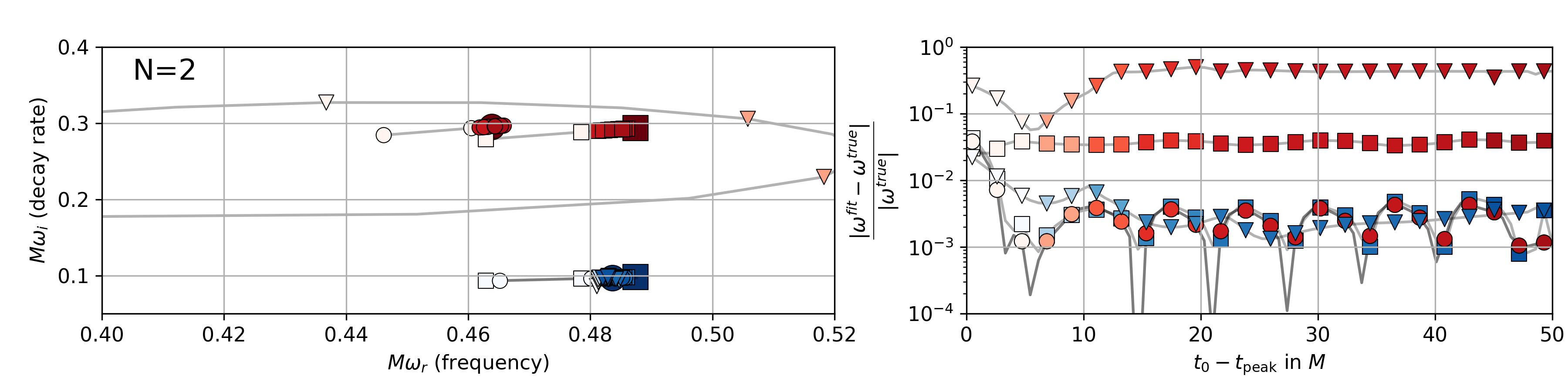}
\includegraphics[width=1.0\linewidth]{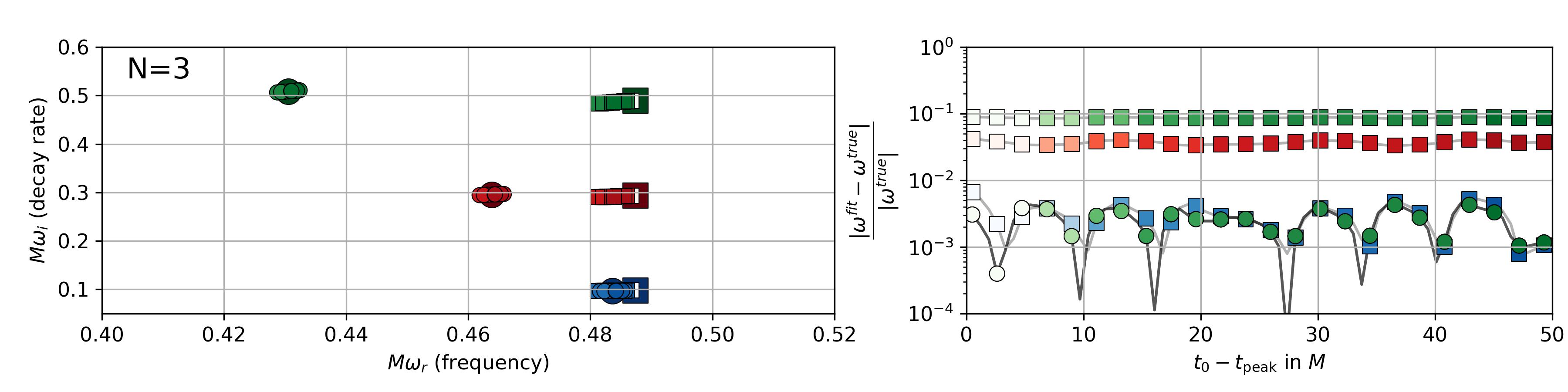}
\caption{Analysis of a GR waveform using different fitting models. This figure has the same structure as Fig.~\ref{PT_qnm}, but for the GR waveform injection.  Note that for the N=2 agnostic fit, the first overtone could not be identified reliably, and so the fit diverges away at late times.}
\label{GR_qnm}
\end{figure*}

We now turn to the question of how well our model fits recover the expected QNM frequencies. The frequencies from a fit are a powerful diagnostic tool: if one can recover modes agnostically, one can check if the recovered spectrum matches a theory predicted one.
To address this question we show the corresponding best-fit results for $t > t_\mathrm{peak}$ for the PT injection in Fig.~\ref{PT_qnm}, and for the GR injection in Fig.~\ref{GR_qnm}. 
In the left panels of each figure we show the PT, GR or AG QNMs obtained using their respective one mode (top), two modes (middle), or three modes (bottom) versions, while the right panels show the convergence of the individual modes in each fit. 
We also show the exact QNMs for each model for $M=1$ for comparison. 
Because the complex QNM frequencies of the PT and GR model are completely determined by one parameter ($M$, which converges at least on percent level), we find comparable and robust convergence for the PT and GR QNMs. 
However, in the AG model, where the QNMs are varied completely independent of their real/imaginary part and mode number, the convergence is much less stable. 

The AG1 models (top panels of Figs.~\ref{PT_qnm} and~\ref{GR_qnm}) explore
a larger part of the plane, and the AG2 models (middle panels of Figs.~\ref{PT_qnm} and~\ref{GR_qnm}) even more so.  For the GR injection, AG2 even exhibits divergent frequencies, see the middle panel of Fig.~\ref{GR_qnm}. 

We remark that the $n=0$ mode frequency of all models agrees very well with the QNM of the injected model.  This is expected, because the fundamental QNM frequency is the most significant feature to fit.  Recovery of the correct fundamental QNM frequency, combined with the fact that this frequency differs by $\sim 1\%$ between the PT and GR potential, explains the 
bias in the underlying mass if the wrong model is used for an injection, as displayed in Figs.~\ref{PT_fig} and \ref{GR_fig}:  The fit arrives at the incorrect mass, to achieve the correct frequency.
This bias in the mass also manifests itself in the small shift of the $n=1$ and $n=2$ QNM frequencies when the wrong model is used to analyze the injection. 
Our conclusion is that the $n=0$ fundamental mode is robustly recovered across fitting models, while their different overtone structures lead to very different fitted overtones.
Note that due to the simple mass dependence of the PT and GR QNMs, their reconstruction is much more constrained for all overtones.  
As a final aspect of our analysis we now turn to the fitted amplitudes and phases. These parameters depend on the initial conditions of our numerical evolutions, and are therefore not as fundamental as the QNM frequencies.

\begin{figure}
\includegraphics[width=1.0\linewidth]{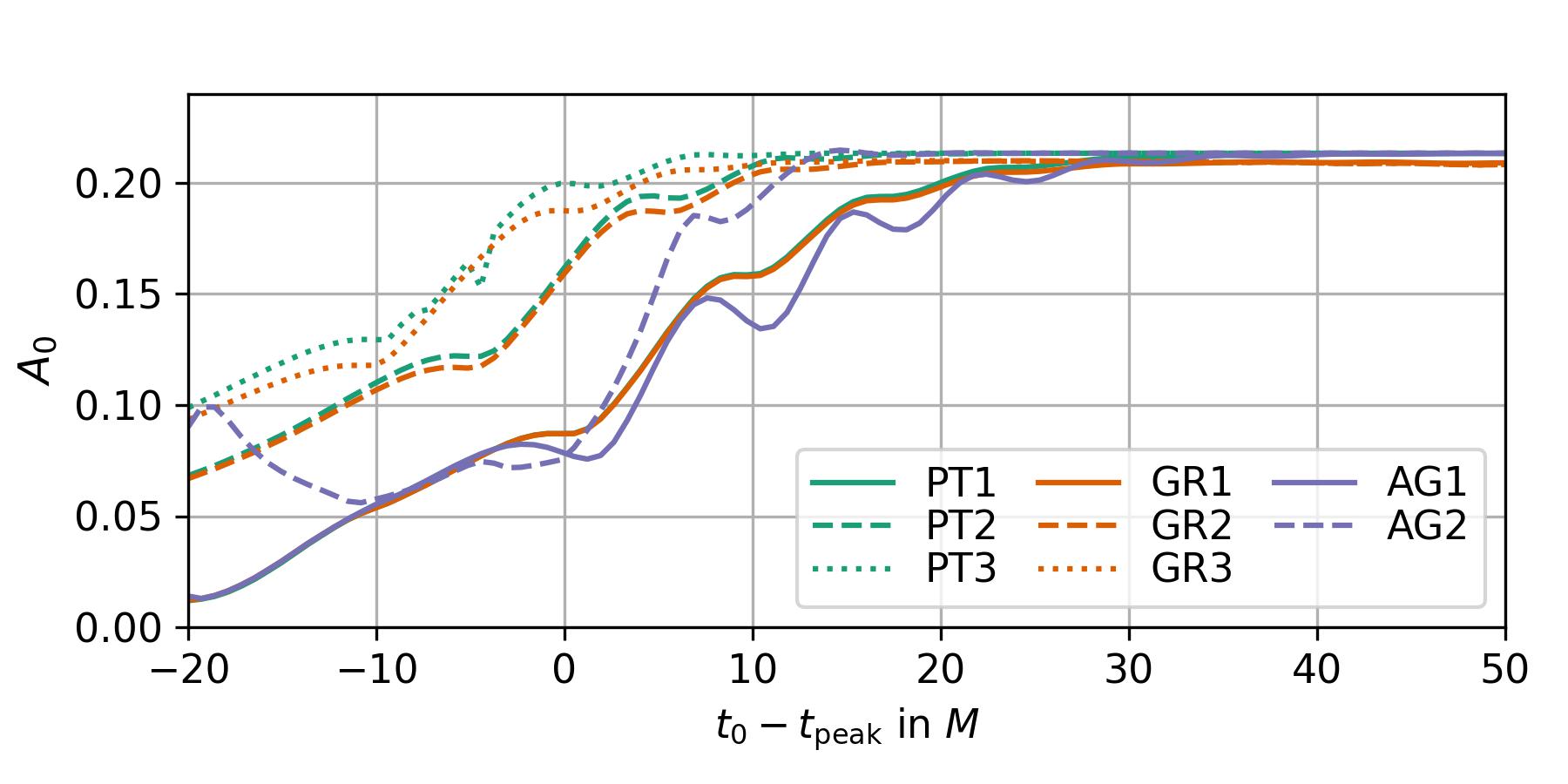}
\includegraphics[width=1.0\linewidth]{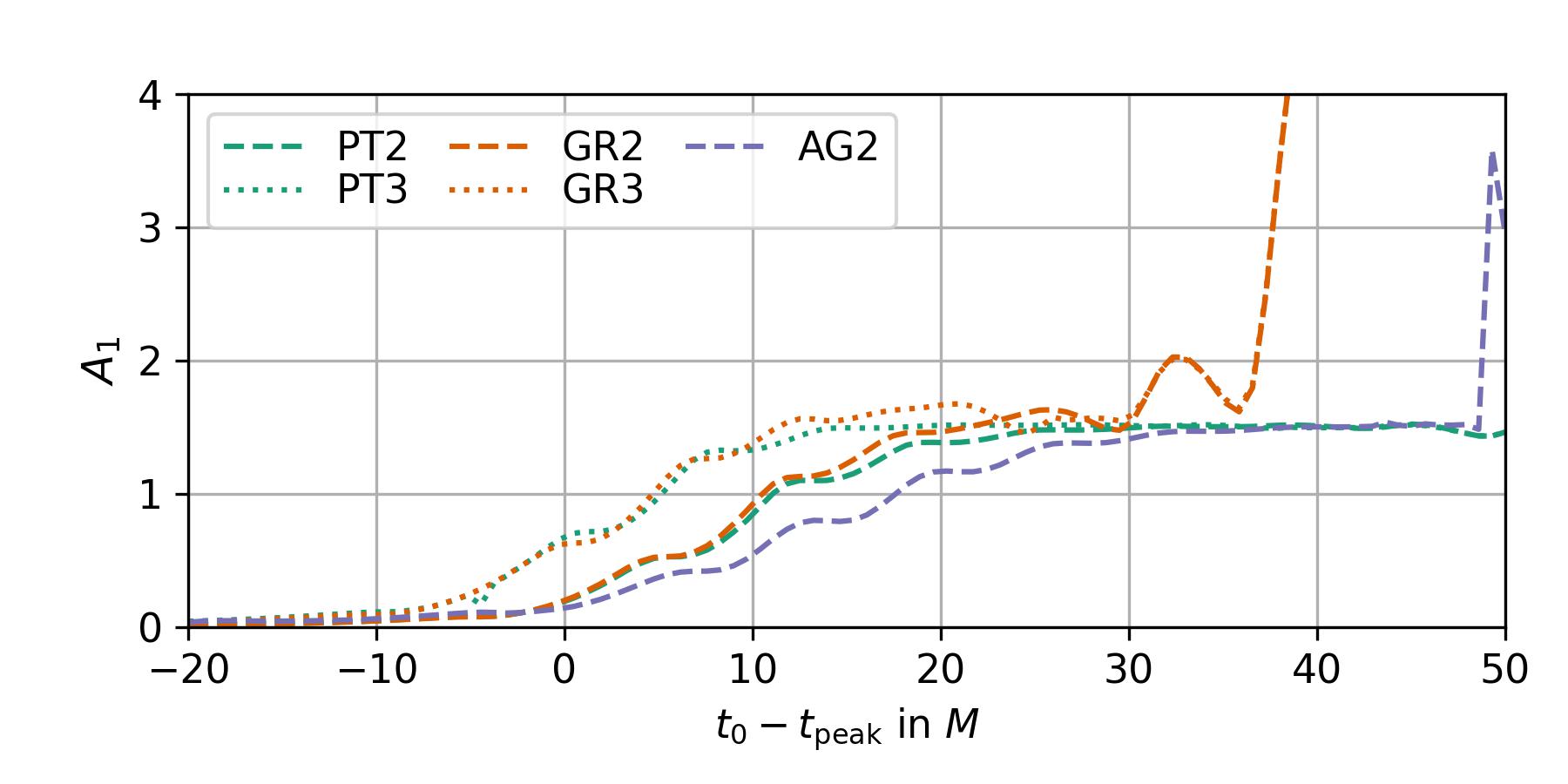}
\caption{
Amplitudes recovered when fitting the PT waveform.  The \textbf{top panel} shows the recovered amplitude of the fundamental mode for fits starting at $t_0-t_{\rm peak}$ for all models.  The \textbf{bottom panel} shows the amplitude of the $n=1$ mode for those models that include this mode.  
}
\label{PT_A}
\end{figure}

\begin{figure}
\includegraphics[width=1.0\linewidth]{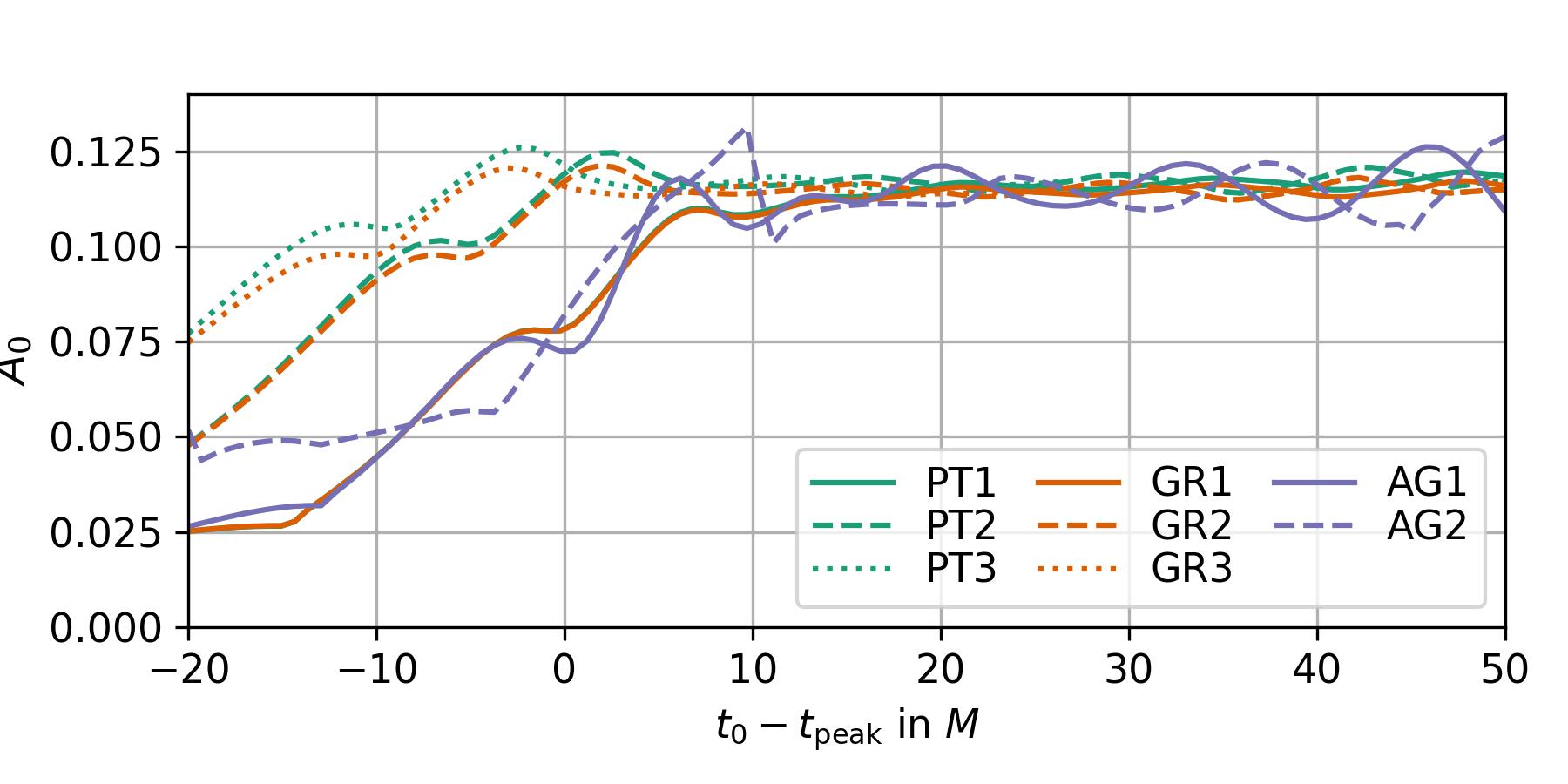}
\includegraphics[width=1.0\linewidth]{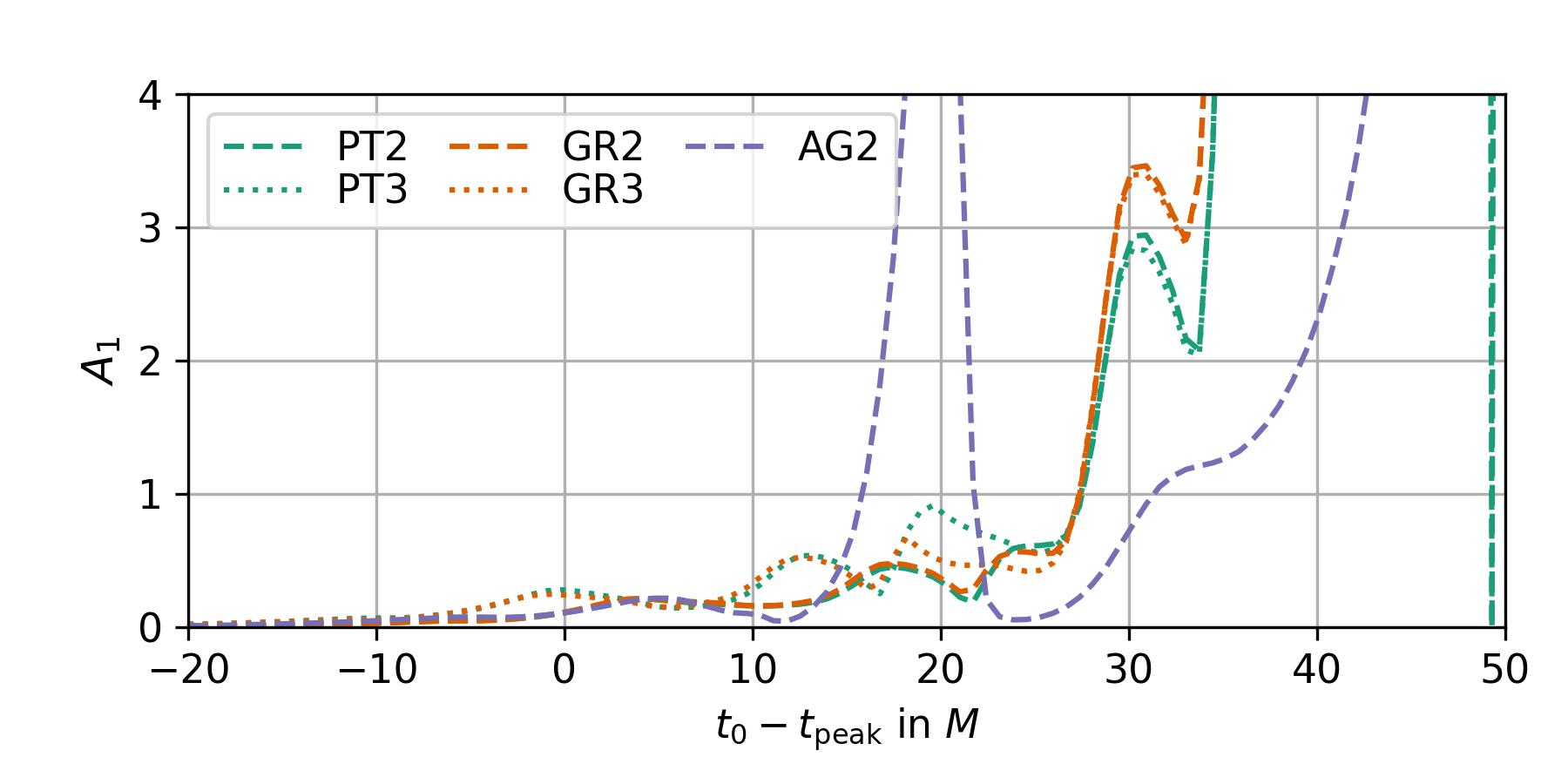}
\caption{Amplitude recovered when fitting the GR waveform.  Data plotted as in Fig.~~\ref{PT_A}.
}
\label{GR_A}
\end{figure}

However, if a given QNM has been clearly found in the signal during fitting, the corresponding amplitude and phase should be constant. 
This should at least be expected qualitatively toward later times, when additional overtones not included in our fits have sufficiently decayed.  In addition, the GR waveform has the added complexity of the presence of the Price tail.
In Figs.~\ref{PT_A} and Fig.~\ref{GR_A} we show the fitted amplitudes $A_0$ and $A_1$ for the fundamental mode ($n=0$, top panels) and the $n=1$ mode (bottom panels) for the PT and GR waveform, respectively. 
For $n=0$ the amplitude can be robustly constrained, within small uncertainties relating to which model has been used. 
For $n=1$ and the PT injection one can also find a stable mode (Fig.~\ref{PT_A}), at least for intermediate times, but much less so for the GR injection (Fig.~\ref{GR_A}).
Further emphasis on the dependence of number of overtones can be seen in the top panels, where the $n=0$ amplitude value is reached at earlier times when more modes are included.

We  note that the amplitudes $A_2$ are extremely unstable and not informative in all cases; therefore, we have not provided them. 
The results for the phases are qualitatively very similar and shown in the Appendix. 
Overall, the findings of nonconstant amplitudes and phases suggest that only
the $n=0$ mode, and potentially the $n=1$ mode for the PT case, can be
robustly inferred from the data. 


\section{Conclusions}\label{con}

It is quite intuitive that adding overtones to a ringdown analysis reduces the mismatch, in particular at earlier times.
It is also clear that a mere reduction in mismatch is not a sufficient criterion to conclude the actual presence of overtones, since introducing a model with additional parameters will generally allow for a better fit. 
However, the observation in Ref.~\cite{Giesler:2019uxc} that it is possible to extract the correct black hole mass even at very early times by using many overtones suggests that it is significant. 
This was carefully investigated in Ref.~\cite{Giesler:2019uxc} by allowing some of the overtone QNMs to deviate from their Kerr prediction, which resulted in finding that the extracted mass was further away from the correct one. 

In our analysis, we have studied a simplified setup that is completely within linear perturbation theory to reduce some of the extra complexities that arise in a full BH ringdown, e.g., when in the ringdown does linear perturbation theory become a good approximation. 
Specifically, we study wave propagation in the GR and P\"oschl-Teller potentials, for which the QNM spectrum is known to very high accuracy and is well understood. 
In both cases, the QNMs can be modeled as a function of only the black hole mass, while the amplitude and phase of each included mode have been treated as free parameters. 
Here, explicit results for amplitudes and phases depend on the choice of the initial data and conclusions need thus to be understood with some caution.

As one part of our analysis, we have recovered the known result (e.g. Ref.~\cite{Baibhav:2017jhs}) that including overtones reduces the mismatch at early times. 
This is not a novel finding but a remarkable reminder that the full nonlinear analysis faces qualitatively similar problems. 
In Refs.~\cite{Ma:2022wpv,Ma:2023cwe} the idea of using additional modes to clean NR signals in the context of a Bayesian analysis has been introduced and successfully demonstrated. 

The unexpected result of our work is that both theory-specific models, scalar Regge-Wheeler and P\"oschl-Teller, perform extremely similarly in the way that they allow for improved estimation of the black hole mass at earlier times. 
Here most strikingly, it is only mildly dependent on the used model to produce the ringdown signal. 
Applying one model to the other signal and vice versa yields very similar results at early and intermediate times, and the ultimate dependence is in the number of overtones included.

Of course, one must be aware that the P\"oschl-Teller approximation gives accurate estimates for the fundamental mode and the imaginary parts of the first few overtones, but the overtones' real parts deviate further with increasing overtone number; see the location of the larger markers in the bottom panels of Fig.~\ref{GR_qnm}. 
The deviations in the QNM frequencies (especially in the fundamental $n=0$ mode) manifest themselves in the fact that using the wrong model at late times results in a plateauing percent level relative error for the mass, while the correct model yields a further decreasing relative error. 

By investigating the robustness of the QNMs obtained by fitting the different models, as well as the fitted amplitudes and phases, we conclude that only the $n=0$ mode can be robustly recovered. 
Already for $n=1$, it is less straightforward to assess the presence of the overtone in the waveform. 
It is likely to be present in the PT waveform, but more difficult to quantify in the GR waveform.

The takeaway message from our analysis, which calls for future studies, should be that including even a rather crude model for the overtones at early times is indeed useful for an earlier extraction of the black hole mass. 
Its limitations only become relevant if the signal can be studied at late enough times, when the asymptotic value gets biased and converges toward a wrong value.  
While the latter can certainly be done within a purely numerical study, contemporary data analysis of real events is limited by moderate signal-to-noise ratio, and it is thus much harder to differentiate between different models. 
Because overtones are particularly sensitive to possible deviations of the Schwarzschild/Kerr space-time or modified dynamics (see, e.g., Refs.~\cite{Cheung:2021bol,Volkel:2022aca,Konoplya:2022pbc}) it is crucial to robustly infer them from future observations.

Because the accuracy of the extracted mass clearly improves at earlier times, even when the wrong model is used, one may ask whether including the overtones is effectively removing the non-QNM contributions related to the initial data. 
In this case, it is questionable whether one can assign a physical significance to higher overtones, or rather understand them as an effective way to improve the parameter estimation by reducing the non-QNM content of the signal originating from the initial data in the linear case or even nonlinear parts in the full problem. 
Indeed, by verifying that the overtones' amplitudes and phases vary as a function of the starting time, one should be convinced that those are not, at least not as anticipated, a real feature in the waveform.  
This, however, does not mean that overtones are not being excited. 
In fact, there is no universal argument why rather generic initial data should not excite them.
Our findings are rather questioning whether standard methods and tests are sufficient to robustly quantify to what extent fitted overtones correspond to physically excited ones.

In the final stage of our work a very comprehensive analysis on overtone fitting has been presented in Ref.~\cite{Baibhav:2023clw}. 
It reports similar findings for some of our main points, although it is not for the P\"oschl-Teller potential. 
Among the different types of ringdown fits that are studied is also a hybrid model, which assumes that some modes are determined by a theory-specific prediction, while some modes are agnostic. 
While our work focuses on the linear case, Ref.~\cite{Baibhav:2023clw} also applies ringdown fitting to numerical relativity waveforms and shows that nonlinear mode effects can become important as well.

\acknowledgments
The authors thank Alessandra Buonanno, Emanuele Berti and Nicola Franchini for useful comments on the manuscript and discussions. The authors thank the anonymous referee for valuable feedback.
S.~H.~V\"olkel acknowledges funding from the Deutsche Forschungsgemeinschaft (DFG) - project number: 386119226. 

\bibliography{literature}
\appendix
\section{Additional material for the phases}\label{app_phase}
\begin{figure}
\includegraphics[width=1.0\linewidth]{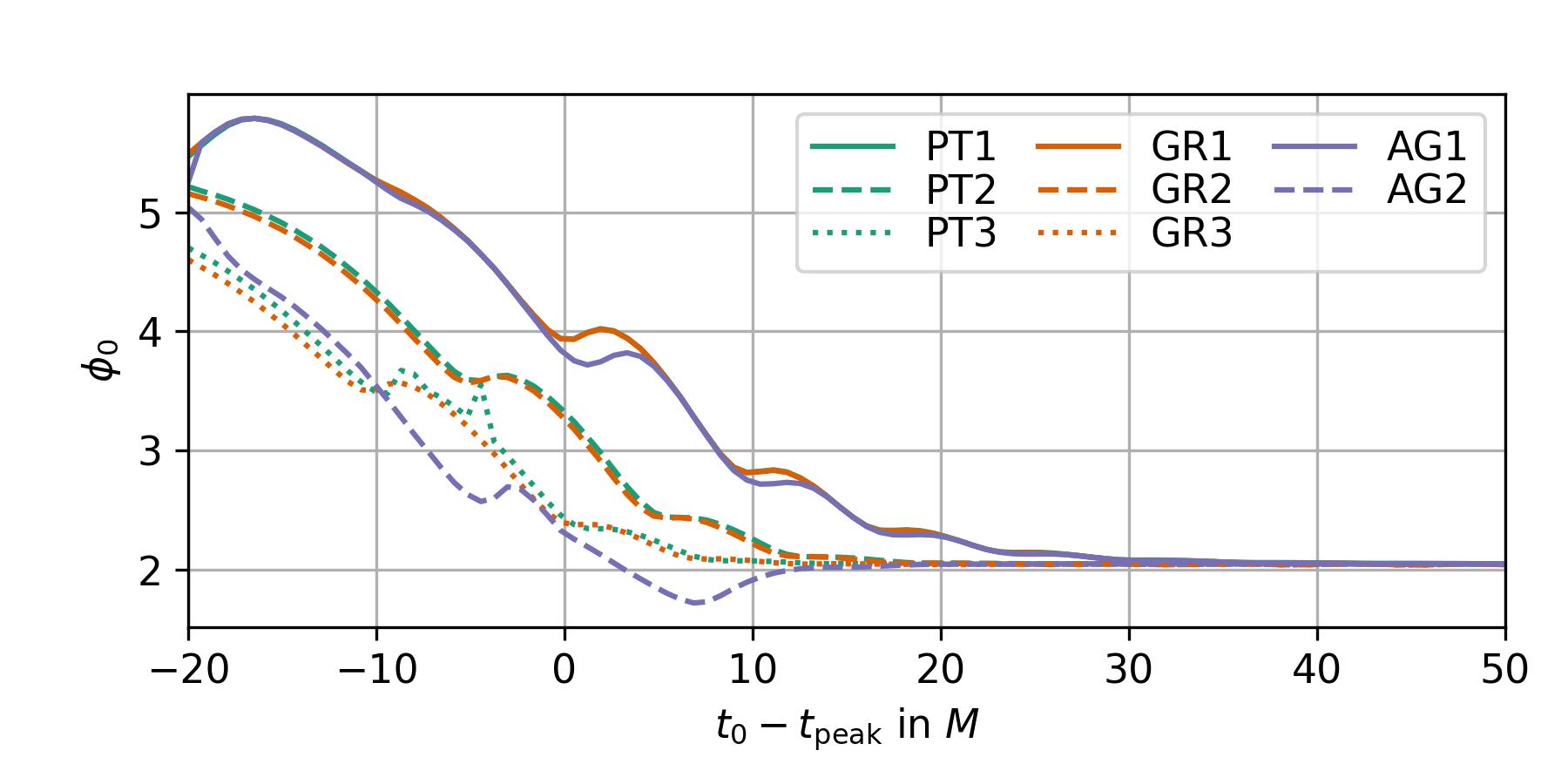}
\includegraphics[width=1.0\linewidth]{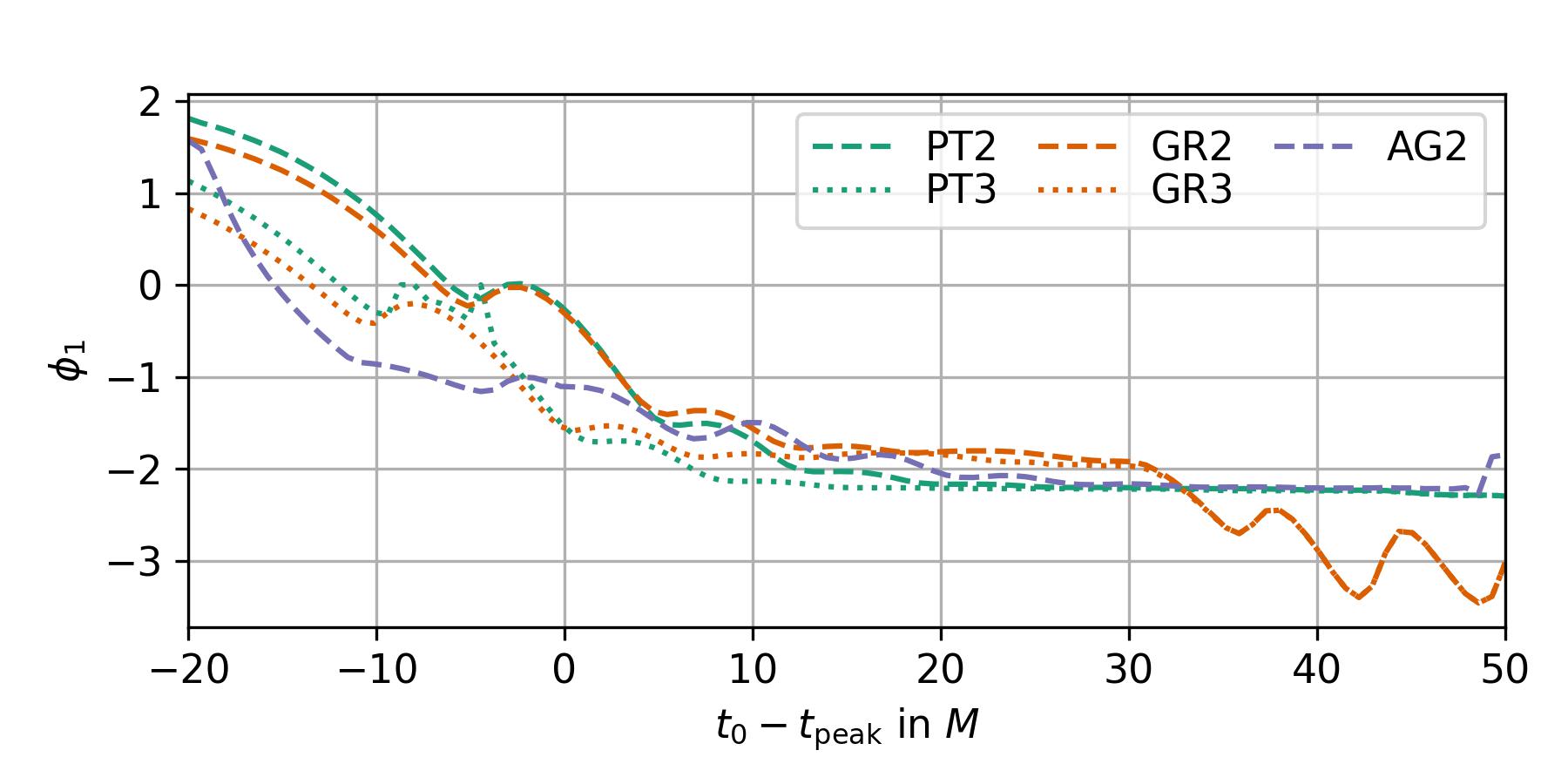}
\caption{Fitted phases of the different models when injecting the PT waveform. 
\textbf{Top:} the $n=0$ mode amplitude $\phi_0$. 
\textbf{Bottom:} The $n=1$ mode amplitude $\phi_1$.
}
\label{PT_phi}
\end{figure}

\begin{figure}
\includegraphics[width=1.0\linewidth]{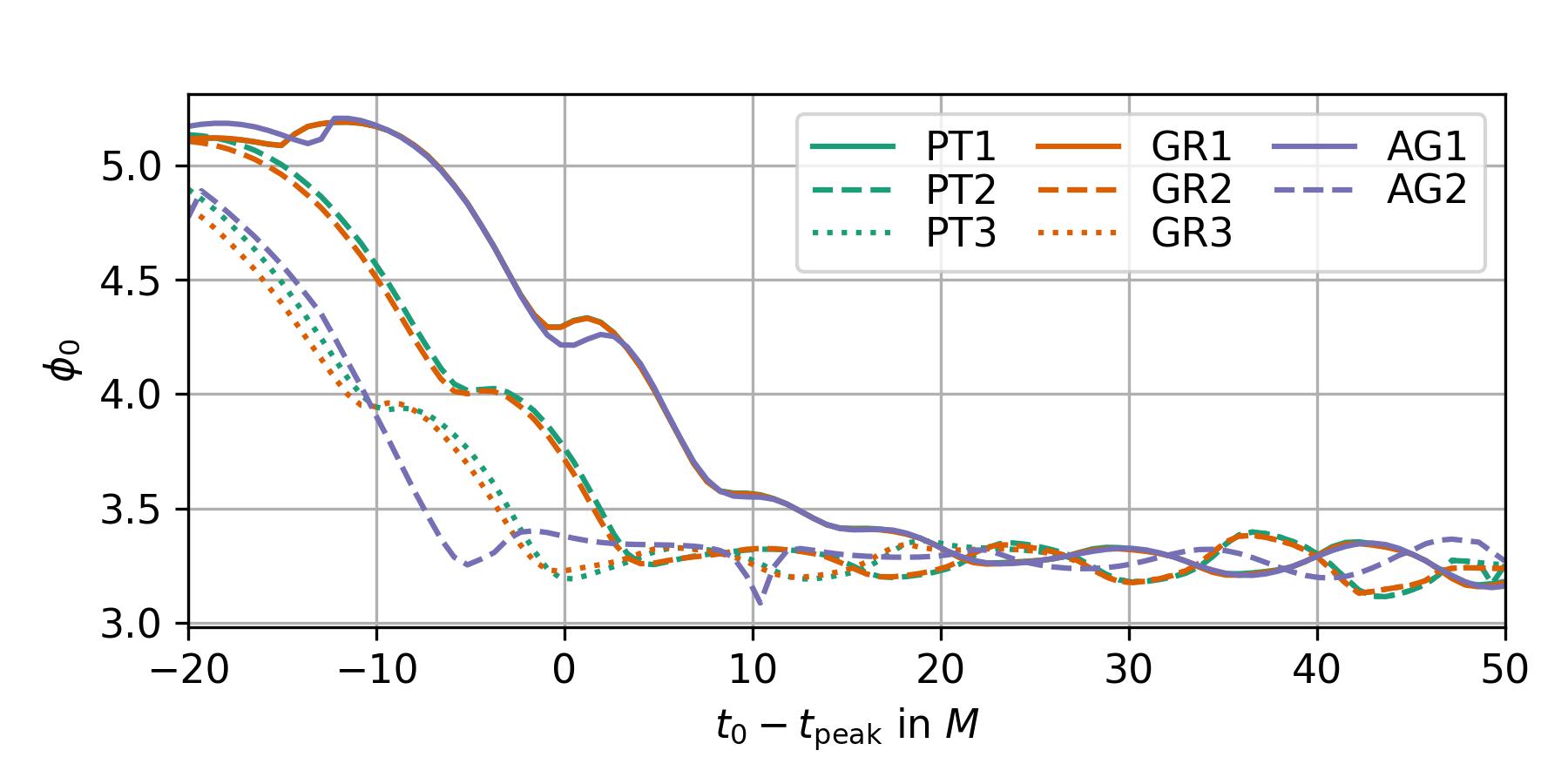}
\includegraphics[width=1.0\linewidth]{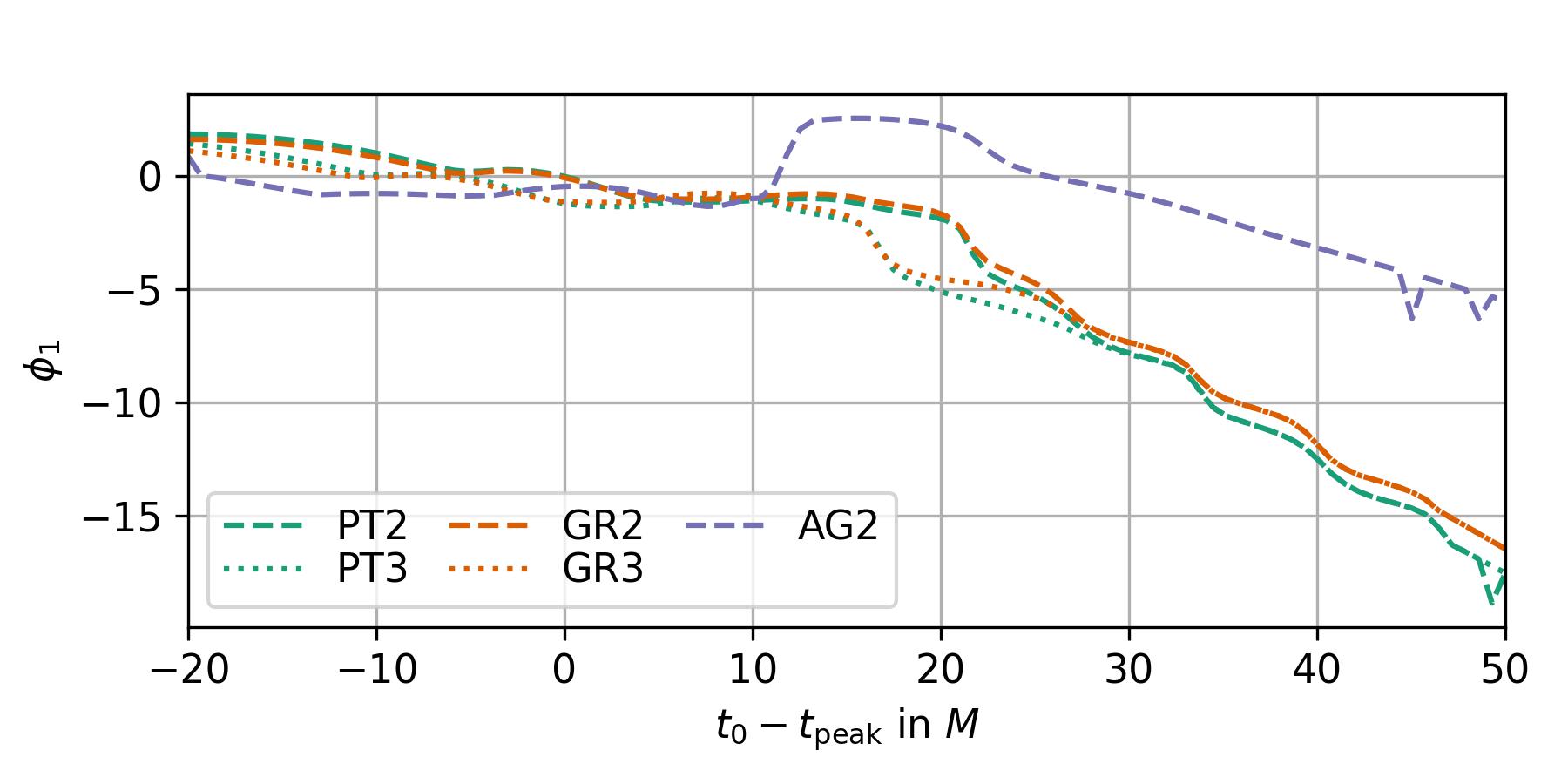}
\caption{The same description as in Fig.~\ref{PT_phi} but for the GR waveform injection.
}
\label{GR_phi}
\end{figure}

Here we report the results for the fitting of the phases for the PT injection in Fig.~\ref{PT_phi} and for the GR injection in Fig.~\ref{GR_phi}. 
Note that during the fitting we consider the range $[0, 2\pi]$; however we have ``unwrapped'' the values here for clearer presentation.
\end{document}